\documentstyle[aps,epsfig]{revtex}
\begin{document}
\begin{flushright}
AS-ITP-98-07 \\
November 1998
\end{flushright}
\vspace*{1cm}

\begin{center}
{\Large  Exclusive Rare Decays of Heavy Baryons to Light Baryons: $\Lambda_b \rightarrow \Lambda \gamma$ and $\Lambda_b \rightarrow \Lambda l^{+} l^{-}$}
\footnotetext{This work was supported in part by the National
Natural Science Foundation of China .}

\vspace*{1cm}
 Chao-Shang Huang and Hua-Gang Yan

\vspace*{0.3cm}

 Institute of Theoretical Physics, Academia Sinica,\\
\hspace*{.3cm} Beijing 100080, China\\
\end{center}
\parindent=0.7cm
\begin{center}
{\large \bf Abstract}\\
\vspace{.5cm}
\begin{minipage}{13cm}
{\small Rare decays  $\Lambda_b \rightarrow \Lambda \gamma$ and $\Lambda_b 
\rightarrow \Lambda l^{+} l^{-}$ (l= e, $\mu$) are examined. We use QCD sum rules to calculate the
hadronic matrix elements governing the decays. The $\Lambda$ polarization in the 
decays is analyzed and it is shown that the polarization parameter in $\Lambda_b
\rightarrow \Lambda \gamma$ does not depends on the values of hadronic form factors. 
And the energy spectrum of $\Lambda$ in $\Lambda_b 
\rightarrow \Lambda l^{+} l^{-}$ is given.
\ 
}
\end{minipage}
\end{center}

\clearpage

\vspace{0.3cm}
{\large \bf I.Introduction}
\vspace{0.3cm}

Processes associated with the flavor-changing neutral current(FCNC) 
$b \rightarrow s$ transition have regained much attention since the 
measurement of FCNC decays of the type $b \rightarrow s \gamma$ by 
CLEO$\cite{many,mannelr}$. It is well known that these processes are forbidden at the tree
level in the Standard Model(SM) and are strongly suppressed by the GIM 
mechanism, particularly for up type quarks. On the other hand. It is very 
sensitive to possible higher mass scales and interactions predicated by
supersymmetic theories, two Higgs doublet models etc.. Such interactions 
shape the $b \rightarrow s$ transition via operators and their Wilson coefficients 
appearing in the low energy $\Delta B=1$ effective Hamiltonian. Hence
the study of such process is an important way to test the CKM sector of the SM
and possibly opens a window to physics beyond the SM.

For the experimental side, there is some data on the exclusive decay 
$B \rightarrow K^{*} \gamma$ and the inclusive decay $B \rightarrow X_s \gamma$
with branching ratio $\cite{data}$
$$ {\rm Br}(B^{+} \rightarrow K^{+*}(892) \gamma ) = (5.7 \pm 3.3) \times 10^{-5},$$
$$ {\rm Br}(B^{0} \rightarrow K^{0*}(892) \gamma )= (4.0 \pm 1.9) \times 10^{-5},$$
$$ {\rm Br}(B \rightarrow X_s \gamma )= (2.3 \pm 0.7) \times 10^{-4}.$$
These data have prompted a number of study aimed at restricting the parameter 
space of various extensions of the SM$\cite{forareview}$. Similar analyses based on 
decays of B mesons have also been performed for the transition $b \rightarrow
s l^{+} l^{-}$, which has not been observed yet$\cite{data}$ and only upper limits on
inclusive branching ratios have been given~\cite{CLEO}:\\
$$ {\rm Br}(b \rightarrow s e^+e^-) <  5.7 \times 10^{-5},~~~~~~ {\rm Br} (b \rightarrow s \mu^+
\mu^-) < 5.8\times 10^{-5}~~~~ {\rm and ~~Br}(b\rightarrow s e^{\pm}\mu^{\mp})< 2.2\times 10^{-5}
 ({\rm at} 90 \% C.L.).$$ 
However, to analyze the
helicity structure of the effective Hamiltonian mediating the transition
$b \rightarrow s$, such analyses are not enough since the information 
on the handedness of the quark is lost in the hadronization process. To access
the helicity of the quarks, analyzing the decay of baryons is the only way.
One experimental drawback of baryon decay compared with B meson decay is that 
the production rate of $\Lambda_b$ baryons in $b$ quark hadronization is
$(10.1_{-3.1}^{+3.9})\%$, which is significantly less than that of B meson$(
Br({\bar b} \rightarrow B^{+})=Br({\bar b} \rightarrow B^{0})=(39.7_{-2.2}^{+1.8})\%, Br(
{\bar b} \rightarrow B_s^{0})=(10.5_{-1.7}^{+1.8})\%) \cite{data}$ , hence the 
relevant analyses have to 
wait for more data on heavy quark decays from future colliders.

For exclusive heavy-to-light decays one has to calculate hadronic matrix elements of 
operators in the effective Hamiltonian between  a heavy hadron and a light hadron, which
is related to the nonperturbative aspect of QCD. There 
are a number of papers to calculate the hadronic matrix elements in exclusive rare 
decays $B\rightarrow K^{(*)}l^+l^-$~\cite{LIU,CFSS}. For exclusive heavy-to-light 
decays $\Lambda_b\rightarrow \Lambda l^+l^-$ there are a lot of form factors to 
describe the hadronic matrix elements. However, for $\Lambda_b$ we may use 
the heavy quark effective theory(HQET). It is well-known that HQET  simplifies greatly the analysis of 
the decay of heavy hadrons in that heavy quark symmetries restrict the number of
from factors, in particular, for the baryonic transition $\Lambda_Q \rightarrow
$ light spin-1/2 baryon, there are only two independent form factors irrelative to
 Dirac matrix of relevant operators$\cite{mannelrr}$; for heavy hadron to
heavy hadron transition, we only have one form factor, which is known as Isgur-Wise 
function $\cite{isgurf}$. The computation of two form factors, $F_1$ and $F_2$, in HQET
 is the main work in analyzing
exclusive decays of $\Lambda_b$ to light baryon. $\Lambda_b \rightarrow \Lambda 
\gamma$ has been in detail investigated by T. Mannel and S. Rocksiegel$\cite{mannelr}$,
where the simple pole model was adopted to
compute $F_1$ and $F_2$. As we know, $\Lambda_b\rightarrow\Lambda l^+ l^-$ has not
been examined yet. The form factors $F_1$ and $F_2$ in $\Lambda_c \rightarrow 
\Lambda$ have been calculated
in nonrelativistic and relativistic quark models~\cite{NRE,RE}. In this paper we employ the 
widely applied approach of QCD sum rules,
which is based on general features of QCD$\cite{qcd}$, to calculate the  $F_1$ and 
$F_2$. For our purpose, we can use some expressions of Ref.~\cite{our} due to the 
similarity. We analyze the $\Lambda$ polarization. An interesting result is that the
polarization parameter in $\Lambda_b\rightarrow \Lambda \gamma$ does not depends on
the values of hadronic from factors, $F_1$ and $F_2$.

This paper is organized as follows. In Sec. II we write down the SM effective Hamiltonian
governing transitions $b \rightarrow s \gamma$ and $b \rightarrow s l^{+} l^{-}$, and give
some information on the corresponding Wilson coefficients. In Sec. III, we compute the
the Form factors $F_1$ and $F_2$, by using QCD sum rules. Sec. IV and Sec. V 
contribute to the analysis of decay $\Lambda_b \rightarrow \Lambda 
\gamma$ and $\Lambda_b \rightarrow \Lambda l^{+} l^{-}$, respectively. In Sec. VI, we
discuss our numerical results along with some relevant points.

\vspace{0.3cm}   
{\large \bf II. Effective Hamiltonian} 
\vspace{0.3cm}   

In the SM, the Hamiltonian relevant for $b \rightarrow s$ transition consists of ten 
operators $O_i, i=1,...,10$. Neglecting terms proportional to $V_{ub} V_{us}^{*}$(the ratio
$|V_{ub}{V_{us}}^{*}/V_{tb}{V_{ts}}^{*}|$ is of order $10^{-2}$), the effective Hamiltonian 
takes the form$\cite{grinsteinsw}$
\begin{equation}
H_W\,=\,4\,{G_F \over \sqrt{2}} V_{tb} V_{ts}^* \sum_{i=1}^{10} C_i(\mu)
O_i(\mu),
\end{equation}
\noindent where 
\begin{eqnarray}
O_1&=&({\bar s}_{L \alpha} \gamma^\mu b_{L \alpha})
      ({\bar c}_{L \beta} \gamma_\mu c_{L \beta}) \nonumber , \\
O_2&=&({\bar s}_{L \alpha} \gamma^\mu b_{L \beta})
      ({\bar c}_{L \beta} \gamma_\mu c_{L \alpha}) \nonumber , \\
O_3&=&({\bar s}_{L \alpha} \gamma^\mu b_{L \alpha})
      [({\bar u}_{L \beta} \gamma_\mu u_{L \beta})+...+
      ({\bar b}_{L \beta} \gamma_\mu b_{L \beta})] \nonumber ,  \\
O_4&=&({\bar s}_{L \alpha} \gamma^\mu b_{L \beta})
      [({\bar u}_{L \beta} \gamma_\mu u_{L \alpha})+...+
      ({\bar b}_{L \beta} \gamma_\mu b_{L \alpha})] \nonumber , \\
O_5&=&({\bar s}_{L \alpha} \gamma^\mu b_{L \alpha})
      [({\bar u}_{R \beta} \gamma_\mu u_{R \beta})+...+
      ({\bar b}_{R \beta} \gamma_\mu b_{R \beta})] \nonumber , \\
O_6&=&({\bar s}_{L \alpha} \gamma^\mu b_{L \beta})
      [({\bar u}_{R \beta} \gamma_\mu u_{R \alpha})+...+
      ({\bar b}_{R \beta} \gamma_\mu b_{R \alpha})] \nonumber , \\
O_7&=&{e \over 16 \pi^2} m_b ({\bar s}_{L \alpha} \sigma^{\mu \nu}
     b_{R \alpha}) F_{\mu \nu} \nonumber , \\
O_8&=&{g_s \over 16 \pi^2} m_b \Big[{\bar s}_{L \alpha} \sigma^{\mu \nu}
      \Big({\lambda^a \over 2}\Big)_{\alpha \beta} b_{R \beta}\Big] \;
      G^a_{\mu \nu} \nonumber , \\
O_9&=&{e^2 \over 16 \pi^2}  ({\bar s}_{L \alpha} \gamma^\mu
     b_{L \alpha}) \; {\bar \ell} \gamma_\mu \ell \nonumber , \\
O_{10}&=&{e^2 \over 16 \pi^2}  ({\bar s}_{L \alpha} \gamma^\mu
     b_{L \alpha}) \; {\bar \ell} \gamma_\mu \gamma_5 \ell. 
\end{eqnarray}
\noindent Here $\alpha,\beta$ are color indices, $b_{R,L}=[(1 {\pm} \gamma_5)/2]b$, 
and $\sigma^{\mu \nu}=(i/2)[\gamma^\mu ,\gamma^\nu ])$; $e$ and $g_s$ are the electronmagnetic and 
the strong coupling constants, respectively. The Wilson coefficients are given in 
Table I~\cite{bm,CFSS}, where NDR denotes
the naive dimensional regularization scheme and ${\overline {MS}}$ denotes the modified minimal subtraction
scheme. The coefficients' dependence on regularization scheme
must disappear in the decay amplitude if all corrections are taken into account. The operator 
responsible for decay $b \rightarrow s \gamma $ is $O_7$, while operators responsible for decay
$b \rightarrow s l^{+} l^{-}$ are $O_7, O_9, O_{10}$. Because $O_7$ does not include
lepton fields, it has to be combined with $\gamma l l $ vertex and lepton fields to
contribute to $b \rightarrow s l^{+} l^{-}$. The result operator is 
$O_7^{'}={e \over {16 \pi^2}}m_b \bar s \sigma_{\mu \nu}(1+\gamma_5)q^{\nu}/q^2 b \bar l \gamma^{\mu} l$.
 
We would like to write $O_9$, $O_{10}$ and $O_7^{'}$ in forms allowing for non-SM couplings(omitting the color indices):
\begin{eqnarray}
O_9&=&{e^2 \over 32 \pi^2}  ({\bar s} \gamma^\mu
     (h_V-h_A \gamma_5) b) \; {\bar \ell} \gamma_\mu \ell , \cr
O_{10}&=&{e^2 \over 32 \pi^2}  ({\bar s} \gamma^\mu
     (h_V-h_A \gamma_5) b) \; {\bar \ell} \gamma_\mu \gamma_5 \ell, \cr 
O_7^{'} &=& {e^2 \over {16 \pi^2}}m_b (\bar s \sigma_{\mu \nu} (g_V -g_A \gamma_5)q^{\nu}/q^2 b) \bar \ell \gamma^{\mu} \ell,
\end{eqnarray}
where $h_V, g_V$ and $h_A, g_A$ are the vector and axial vector couplings respectively, in order 
to discuss possible effects of models beyond SM. In general the parameters $h_V$ and $h_A$ in $O_{10}$ may be
different
from those in $O_9$ and for the sake of simplicity we consider the case of same parameters as shown in Eq.(3). In the
SM
$h_V=1, h_A=1, g_V=1, g_A=-1$ if we neglect the mass of strange quark in respect to b quark. 

As mentioned in Sec.I, the heavy quark symmetries restrict the number of form factors 
to two~\cite{mannelrr}:
\begin{equation}
<\Lambda(p,s)|\bar s \Gamma b| \Lambda_b (v,s^{'})> = \bar u_\Lambda (p,s)\{F_1 (p.v)+ \not\! v F_2 (p.v) \} \Gamma u_{\Lambda_b} (v,s^{'})
\end{equation}
The following section contributes to the computation of $F_1$ and $F_2$ by QCD sum rules in HQET.

\vspace{0.3cm}
{\large \bf III. Computation of $F_1$ and $F_2$}
\vspace{0.3cm}

In Ref.\cite{our}, using QCD sum rules, we computed the form factors of $\Lambda_b$ to 
$p$ transition. Here we follow the approach and use the some expressions obtained in 
the work.

To compute $F_1$ and $F_2$ within the QCD sum rule approach we need to consider the three-point
correlator
\begin{equation}
\Pi (P^{'}, P, z)=i^2\int d^4xd^4ye^{ik\cdot x-iP\cdot y}
<0|T\tilde{j}^{v}(x)\bar{h}_{v}(0)\Gamma s(0)\bar{j}(y)|0>~,
\end{equation}
of the flavor-changing current $\bar h_v \Gamma s$ and of $\Lambda_b$ current $\tilde
{j}^v$ and $\Lambda$ current $j$, where $P^{'}=m_b v + k$ and $z= P\cdot v$.
The baryonic currents for $\Lambda_b$ are
$$ \tilde{j}^v=\epsilon^{abc}(q_1^{{\rm T}a}C\tilde{\Gamma}\tau q^b_2)h_v^c~, $$
where $\tilde \Gamma$ has too kind of choice, $\gamma_5$ and $\gamma_5 \not\! v$. We
choose $\gamma_5$ for the sake of simplicity since the numerical differences resulting from
the different choices of $\tilde {j}^{v}$ are not significant$\cite{daihll}$.
 
The three-quark $tensor\ currents$ for proton and $\Lambda$ are~\cite{sn}
$$J_p^T(x_1,x_2,x_3)=\sigma^{\mu \nu} \gamma_5 d^a (x_1)u^b(x_2) C \sigma_{\mu \nu} u^c(x_3) \epsilon^{abc}, $$
$$J_{\Lambda}^T(x_1,x_2,x_3)=\sigma^{\mu \nu} \gamma_5 d^a (x_1)u^b(x_2) C \sigma_{\mu \nu} s^c(x_3) \epsilon^{abc}. $$
After Fierz transformation, they can be written in $S+P$ form
\begin{eqnarray}
J_p^T(x_1,x_2,x_3)&=&4[u^a(x_3) u^b(x_2) C \gamma_5 d^c(x_1)+\gamma_5 u^a(x_3) u^b(x_2) C d^c(x_1)] \epsilon^{abc},\cr 
J_{\Lambda}^T(x_1,x_2,x_3)&=&4[s^a(x_3) u^b(x_2) C \gamma_5 d^c(x_1)+\gamma_5 s^a(x_3) u^b(x_2) C d^c(x_1)] \epsilon^{abc}.
\end{eqnarray}
The similarity allows us to use directly the analytical expressions obtained in the 
sum rule analysis of $p$ final state to compute the form factors of $\Lambda$ final 
state if we neglect the mass of s quark.

After inserting a complete set of physical intermediate states,
as the phenomenological consequence of (5), we have
\begin{equation}
\Pi(P^{'}, P, z)= f_{\Lambda_b} f_\Lambda {2 \over {(\omega - 2 \bar \Lambda)}}
        P_{+} \Gamma [F_1(z)+F_2(z) \not\! v ] {{\not\! P + m_\Lambda}\over{P^2-m
_\Lambda^2}} + {\rm res},
\end{equation}
where $P_{+}=(1+\not\! v)/2$, $\bar{\Lambda}=m_{\Lambda_b}-m_b$, $\omega=2k
 \cdot v$ and $f_{\Lambda_b}$, $f_\Lambda$ are the so-called
"decay constants" which are defined by
\begin{equation}
<0|\tilde{j}^v|\Lambda_b >=f_{\Lambda_b}u~, \\
<p(P)|\bar j|0>=f_\Lambda \bar u(P).
\end{equation}
They can be found in Ref.$\cite{daihll,hwangy}$ and Refs.$\cite{sn,nrl}$, respectively. To obtain
(7) we have taken into account (4) and the heavy quark limit.

By introducing the assumption of quark-hadron duality, the contribution
from higher resonant and continuum states can be treated as
\begin{equation}
{\rm res}=\int_{D^{'}} d\nu ds{{\rho_{pert}(\nu,s,z)} \over{(\nu-\omega)(s-P^2)}
}.
\end{equation}
The region $D^{'}$ is characterized by one or two continuum thresholds
$\nu_c$, $s_c$.
From the theoretical point of view, $\Pi(P^{'}, P, z)$ is combination of the perturbative
contribution and the condensate contribution. In order to incorporate the above assumption,
we should express the perturbative term in the form of dispersion relation
\begin{equation}
\Pi_{pert}^i (\omega,P^2,z)=\int d\nu ds{{\rho^i (\nu,s,z)}\over{(\nu-\omega)(s-P^2)}}, ~~~ (i=1,2),
\end{equation}
where $i=1,2$ denote the different
terms associated with $F_1$ and $F_2$, respectively. In the standard way, we employ a double 
Borel Transformation $\omega\rightarrow M, P^2\rightarrow T$ in order to suppress the higher 
excited state and continuum state contributions. Because the $SU_3$ violation effects
are small in QCD sum rule analyses of three point functions for mesons~\cite{hl}, we
expect the effects are probably even smaller for baryons. Therefore, we neglect the s
quark mass in calculations and quote from Ref$\cite{our}$ the analytical expressions
of $\rho^i (\nu,s,z), i=1,2$ and the condensate contributions. Thus, the resulting
Borel transformed sum rules for $F_1$ and $F_2$ can be written as 
\begin{equation}
\begin{array}{lll}
- 2f_{\Lambda_b}f_{\Lambda} F_1 e^{-2\bar{\Lambda}/M-m^2_{\Lambda}/T} &=&\int_0^{\nu_c}d\nu \int_{0}^{2\nu
z}ds \rho_{pert}^1 e^{-s/T-\nu/M}
- {1\over 3}\langle \bar q q \rangle^2 - \cr
& &{1\over {32\pi^4}}\langle\alpha_sGG\rangle \int_0^{T/4} (1-{{4\beta}\over T})  e^{-4\beta(
1-4\beta/T)/M^2-8\beta z/(TM)}d\beta,\cr
- 2f_{\Lambda_b}f_{\Lambda} m_{\Lambda} F_2 e^{-2\bar{\Lambda}/M-m^2_{\Lambda}/T} &=&\int_0^{\nu_c}d\nu \int_{0}^{2
\nu z}ds \rho_{pert}^2 e^{-s/T-\nu/M} + \cr
& & {1\over {8\pi^4}}\langle\alpha_sGG\rangle \int_0^{T/4}(1-{{4\beta}\over T}){\beta \over M
}e^{-4\beta(1-4\beta/T)/M^2-8\beta z/(TM)}d\beta,
\end{array}
\end{equation}
and 
\begin{equation}
\begin{array}{lll}
\rho_{pert}^1&=&{1\over{32\pi^4\sigma^3}}[-2z^3\sigma^3-(-s+z(\nu+2z))^3+
3z^2(-s+z(\nu+2z))\sigma^2], \cr
\rho_{pert}^2&=&{-1\over{64\pi^4\sigma^3}}[s-2z^2+z(-\nu+\sigma)]^2[\nu s+8z^3-
4z^2(-2\nu+\sigma)-2z(-\nu^2+5s+\nu\sigma)],
\end{array}
\end{equation}
with $\sigma=\sqrt{-4s+(\nu+2z)^2}$. It should be mentioned here that we only retain the condensates
with dimension lower than 7. In the numerical analysis, the "decay constants" and some other 
constants we used are$\cite{daihll,hwangy,nrl,data}$:
\begin{equation}
\begin{array}{rcl}
& &m_{\Lambda_b}=5.64 {\rm GeV}, m_\Lambda=1.116 {\rm GeV}, f_{\Lambda_b}=\sqrt{0.0003} {\rm GeV}^3\cr
& &f_\Lambda=0.0208 {\rm GeV}^3, \bar \Lambda=0.79 {\rm GeV}, \langle\bar{q}q\rangle \simeq-(0.23~ {\rm GeV})^3~,\cr
& &\langle\alpha_sGG\rangle \simeq0.04~ {\rm GeV}^4~.
\end{array}
\end{equation}

Again, owing to the small difference between $p$ and $\Lambda$ both in their mass and "decay constant",
the results here are similar to those of Ref.$\cite{our}$. With the threshold $\nu_c=3.5$GeV,
we can have a reasonably good window for $F_1$, where $1.5{\rm GeV}<{{4T}\over{m_b}}
=M<1.9{\rm GeV}$. The results are given in Fig.1 and Fig.2 respectively, where the different 
curves correspond to different choices of the Borel parameters. The comments made in Ref.$\cite{our}$
concerning about the sum rules, such as the condensates dominance property(for about $55\%$)
, still hold here.  One can see from figures 1 and 2 that $R=F_2/F_1=-0.42$ at 
$q^2=q^2_{max}$ which is almost the same as that of p final state.

\vspace{0.3cm}
{\large \bf V. Decay $\Lambda_{b} \rightarrow \Lambda \gamma$}
\vspace{0.3cm}

As analyzed in Ref.~\cite{mannelr}, the long distance
contribution to $\Lambda_b\rightarrow \Lambda \gamma$ is negligibly small, so we shall only consider the 
short distance contribution to the decay.
The matrix element of the operator $O_7$ between the initial and final state is
\begin{equation}
 \langle \Lambda(p,s), \gamma(k,\varepsilon)| O_7 | \Lambda_b(v,s') \rangle = {e\over 32\pi^2} m_b \langle \Lambda(p,s) |\bar s \sigma_{\mu\nu}
(g_V-g_A\gamma_5) b | \Lambda_b(v,s') \rangle
\langle \gamma(k,\varepsilon)|F^{\mu\nu}|0 \rangle .
\end{equation}
In order to study the helicity of the final quark, we are concentrated on the decay rate
of unpolarized $\Lambda_b$ baryons into light baryons with a definite spin directions. 
We obtain  the decay rate from Eqs.(4) and (14) as follows:
\begin{eqnarray}
\Gamma &=&
{ C_7^2 G_F^2 \over \pi} (V_{ts}^*V_{tb})^2
({e\over 16\pi^2})^2 m_b^2 m_{\Lambda_b}^3 (1- x^2) \{ {g_V^2 + g_A^2 \over 2} [(1-2x^2+x^4 )|F_1|^2 + 2(x-2x^3+x^5)F_1 F_2 \cr
&& + (x^2-2x^4+x^6)|F_2|^2 ] +g_V g_A (v.s) [ (2x-2x^3) |F_1|^2 + 2(2x^2-2x^4)F_1 F_2 + (2x^3-2x^5)|F_2|^2 ] \}
\end{eqnarray}
where $x=m_\Lambda/m_{\Lambda_b}$. Our result is exactly the same as that of 
Ref.$\cite{mannelr}$ with real $F_1$ and $F_2$, as it should be. In order to compare 
the result with experiment we rewrite the rate in terms of the polarization variables as 
defined in Ref.$\cite{mannelr}$
\begin{equation}
\Gamma = \Gamma_0 ~ [1+ \alpha \bf \hat p \cdot S_\Lambda]
\end{equation}
where ${\bf \hat p}$ is the unit momentum vector of the $\Lambda$ and $S_\Lambda$ is its
spin vector. 
From expression (15), it is straightforward to find that 
\begin{eqnarray}
\alpha &=& {2x \over (1+x^2)} {2g_V g_A \over (g_V^2 + g_A^2)}, \\
\Gamma_0 &=& { C_7^2 G_F^2 \over 2\pi} (V_{ts}^*V_{tb})^2 ({e\over 16\pi^2})^2 m_b^2 m_{\Lambda_b}^3 (1- x^2)^3 (F_1+F_2 x)^2 (g_V^2+g_A^2).
\end{eqnarray}
Eq. (17) is different from that in Ref.~\cite{mannelr} and means that the polarization
parameter $\alpha$ is independent of the form factors $F_1$ and $F_2$. The conclusion
that $\alpha$ does not depend on the hadronic structure has also been obtained in Ref.
~\cite{chh}. However, in Ref.~\cite{chh} $F_2$ = 0 has been assumed so that the
conclusion is trivially obtained and the factor 2x/(1+$x^2$) in Eq. (17) is missed
there. The QCD sum rules analyses give that $F_1 =0.50\pm 0.03$ and $F_2 =-0.10\pm 0.03$ 
at the point $p^0=({m_{\Lambda_b}}^2+{m_\Lambda}^2)/(2m_{\Lambda_b})=2.93 \rm GeV$. 
Although this point is to the disadvantage of HQET application, which justifies Eq.(4), for
the recoil at this point is the largest, we still assume that the heavy quark symmetries are 
applicable to some extent at this point. Taking $x=0.20$, we have 
\begin{equation}
 \alpha = 0.38 \cdot {2 g_V g_A \over (g_V^2+g_A^2)},
\end{equation}
and\\
\begin{equation}
\Gamma_0 = (1.06\pm 0.16) \times 10^{-17}(g_V^2+g_A^2){\rm GeV},
\end{equation}
where the uncertainty is rooted in the uncertainties of $F_1$ and $F_2$. In the SM, we take $g_V=1$, $g_A=-1$, the corresponding branching ratio is
\begin{equation}
 {\rm BR}(\Lambda_b \rightarrow \Lambda \gamma)=(3.7\pm 0.5) \times 10^{-5}
\end{equation}
which is within the range obtained in Ref. $\cite{mannelr}$.

\vspace{0.3cm}
{\large \bf V. Decay $\Lambda_{b} \rightarrow \Lambda l^{+} l^{-}$}
\vspace{0.3cm}

The decay of lepton final state is a little more involved than that of $\gamma$ final 
state in the integration over phase space. The relevant matrix element of the process are:
\begin{eqnarray}
\langle \Lambda(p,s), l^{-}(p_1,s_1),l^{+}(p_2,s_2)|O_7^{'}| \Lambda_b(v,s') \rangle &=& {e^2\over 16\pi^2} m_b \langle \Lambda(p,s) |\bar s \sigma_{\mu\nu}
(g_V-g_A\gamma_5){q^{\nu}\over q^2} b | \Lambda_b(v,s') \rangle \cr
&&\langle l^{-}(p_1,s_1),l^{+}(p_2,s_2)|\bar \ell \gamma^\mu \ell|0 \rangle  \cr
\langle \Lambda(p,s), l^{-}(p_1,s_1),l^{+}(p_2,s_2)|O_9| \Lambda_b(v,s') \rangle &=& {e^2\over 32\pi^2}\langle \Lambda(p,s) |\bar s \gamma_\mu (h_V-h_A \gamma_5) b | \Lambda_b(v,s') \rangle \cr
&&\langle l^{-}(p_1,s_1),l^{+}(p_2,s_2)|\bar \ell \gamma^\mu \ell|0 \rangle  \cr
\langle \Lambda(p,s), l^{-}(p_1,s_1),l^{+}(p_2,s_2)|O_{10}| \Lambda_b(v,s') \rangle &=& {e^2\over 32\pi^2}\langle \Lambda(p,s) |\bar s \gamma_\mu (h_V-h_A \gamma_5) b | \Lambda_b(v,s') \rangle \cr
&&\langle l^{-}(p_1,s_1),l^{+}(p_2,s_2)|\bar \ell \gamma^\mu \gamma^5 \ell|0 \rangle 
\end{eqnarray}
To obtain from the above expressions the differential width with respect to the
energy $E$ of $\Lambda$ is a matter of algebra, we obtain
\begin{equation}
{d\Gamma \over {dy}} = A(y) + s\cdot v B(y) 
\end{equation}
where $y=E/m_{\Lambda_b}$. The expressions of $A(y)$ and $B(y)$ are given in the appendix for they 
are somewhat lengthy. In the expressions, we have replaced $C_9$ with $C_9^{eff}$, which
takes the main long distance contribution into consideration by following formalism 
$\cite{grinsteinsw,lms,dt}$:
\begin{equation}
C_9^{eff}=C_9+(3C_1 + C_2) [h(x,s)+k \sum\limits_{i=1}^{2} {\pi \Gamma(\psi \rightarrow
l^{+} l^{-}) M_{\psi_i} \over {q^2-M_{\psi_i}^2+iM_{\psi_i} \Gamma_{\psi_i}}}],
\end{equation}
where
$$h(x,s)=-[{4 \over 9}{\rm ln}x^2-{8 \over {27}}-{{16} \over 9}{x^2 \over s}+{4 \over 9}
\sqrt{ {{4 x^2} \over s}-1} (2+{{4 x^2} \over s}){\rm arctan}({{4x^2}\over s} -1 )^{-1/2}]$$
if $s<4 x^2$ and
$$h(x,s)=-\{ {4 \over 9}{\rm ln}x^2-{8 \over {27}}-{{16} \over 9}{x^2 \over s}+{2 \over 9}
\sqrt{1- {{4 x^2} \over s}} (2+{{4 x^2} \over s})[{\rm ln}|{{1+\sqrt{1-4x^2/s}} \over {1-\sqrt{
1-4x^2/s}}}| - i \pi ] \}$$
if $s>4 x^2$, with $x=m_c/m_b$ and $s=q^2/m_b^2(q^2=M_{l^{+} l^{-}}^2)$.
As in Sec. V, 
we write the total width in the form of Eq.(16)
\begin{equation}
\Gamma^{'} = \Gamma_0^{'} ~ [1+ \alpha^{'} \bf \hat p\cdot S_\Lambda].
\end{equation}
Our calculation yields
\begin{eqnarray}
\Gamma_0^{'} &=& (0.0045 g_A^2 + 0.0045 g_V^2 + 0.0075 g_A h_A + 1.64 h_A^2
   - 0.0069 g_V h_V + 1.64 h_V^2 )\cdot (1.00\pm 0.24) \times 10^{-17} {\rm GeV} \cr
\alpha^{'} &=& {{0.0037 g_V h_A  + 0.0037 g_A g_V - 0.037 g_A h_V - 1.73 h_A h_V
 }\over {
0.0045 g_A^2 + 0.0045 g_V^2 + 0.0075 g_A h_A + 1.64 h_A^2
   - 0.0069 g_V h_V + 1.64 h_V^2}} \times (1.00\pm 0.08) ,
\end{eqnarray}
where the uncertainties are mainly rooted in the uncertainties of $F_1$ and $F_2$. In the SM, we take $h_V=1, h_A=1, 
g_V=1$, $g_A=-1$, then $2\Gamma_0^{'}$ gives the total decay 
width $(6.57\pm 1.58)\times 10^{-17}$GeV and $\alpha^{'}= - 0.54\pm 0.04 $ . 
The differential widths, $A(y)$ and ${\sqrt{y^2-x^2} \over {y}} B(y)$ (the spin-dependent term of the energy spectrum in the rest frame of $\Lambda_b$), are given 
in Fig. 3 and Fig. 4 respectively.

\vspace{0.3cm}
{\large \bf VI. Discussions}
\vspace{0.3cm}

 In this paper we have analyzed some features of the rare decays $\Lambda_b \rightarrow \Lambda \gamma$
and $\Lambda_b \rightarrow \Lambda l^{+} l^{-}$, using an approach based on three-point function QCD 
sum rules to compute the relevant form factors.

 We have considered only the short distance contribution for $\Lambda_b\rightarrow
\Lambda \gamma$. There are some estimations on
the long distance contribution in Ref.$\cite{mannelr}$, the results turn out to be 
negligibly small, the
decay $\Lambda_b \rightarrow \Lambda \gamma$ is dominated by the short distance piece.
For $\Lambda_b\rightarrow\Lambda l^+ l^-$ we have taken into account
the main long distance contributions included in the coefficient $C_9^{eff}$.   

 As we have mentioned in Sec. IV, in applying QCD sum rules analysis for 
$\Lambda_b \rightarrow \Lambda \gamma$, we need to assume the the heavy quark symmetry
is applicable at its most disadvantage phase space corner. For the process 
$\Lambda_b \rightarrow \Lambda l^{+} l^{-}$, the heavy quark symmetry works well in most 
of phase space. In particular, one can see from the figure 1 that the form factor $F_1$
 becomes larger when z approaches the zero recoil point so that its contributions to
the decay are dominant.  

Our results show that the total decay width $\Gamma(\Lambda_b\rightarrow\Lambda l^+l^-)$
is larger than that for $\Lambda_b\rightarrow\gamma$, which is due to the dominance of
the long distance (resonance) contributions to the decay $\Lambda_b\rightarrow\Lambda 
l^+l^-$. From the expressions of Eq.(19) and Eq.(26), in addition to that the total 
width of lepton final state is greater than that of $\gamma$ final state, its 
polarization effect is more remarkable than that of
$\gamma$ final state either. So the measurements of polarization
parameter in the decay  $\Lambda_b \rightarrow 
\Lambda l^{+} l^{-}$ is of a complement to those in  $\Lambda_b\rightarrow\Lambda\gamma$ in order to discover new physics
.  From Eq.(26), we can also find that the contributions to decay
$\Lambda_b \rightarrow \Lambda l^{+} l^{-}$ mainly come from $O_9$ and $O_{10}$.

We emphasize that the polarization parameter in $\Lambda_b\rightarrow\Lambda\gamma$ is
independent of the hadronic structure of $\Lambda$ in the heavy quark limit so that it is a good quantity to
probe new physics beyond SM. In addition to non-SM couplings in $o_7$, we have also considered the non-SM couplings
 in $O_9$ and $O_{10}$ and examined their effects in rare decays of $\Lambda_b$. It is expected that the measurements
 of $\Lambda_b\rightarrow \Lambda \gamma$ and $\Lambda_b\rightarrow\Lambda l^+l^-$ will give us some information on
 physics beyond the SM.

\clearpage
{\large \bf Appendix: Expression of $A(y)$ and $B(y)$}
\begin{eqnarray}
A(y)&=&\frac{1}{3}{{{k_0}}^2}m_{\Lambda_b}^3\{ {{C_7}^2}{{m_b}^2}r
     \{ {{F_1}^2} [ {{g_A}^2}
           ( 4 - 8{x^2} + 4{x^4} )  +
          {{g_V}^2}( 4 - 8{x^2} + 4{x^4} )
          ]  + {F_1}{F_2} [ {{g_A}^2}
           ( 8x - 16{x^3} + 8{x^5} )  + \cr
&&          {{g_V}^2}( 8x - 16{x^3} + 8{x^5} )
           ]  + {{F_2}^2} [ {{g_A}^2}
           ( 4{x^2} - 8{x^4} + 4{x^6} )  +
          {{g_V}^2}( 4{x^2} - 8{x^4} + 4{x^6} )
           ]  )  + {{C_7}^2}{{m_b}^2}
     ( {{F_1}^2} [\cr
&& {{g_V}^2}
           ( -4 - 12x - 4{x^2} + 4y )  +
          {{g_A}^2}( -4 + 12x - 4{x^2} + 4y )
           ]  + {F_1}{F_2}
        [ {{g_V}^2}
           ( 16x - 8{x^3} - 24y - 16xy ) \cr
&& +
          {{g_A}^2}
           ( 16x - 8{x^3} + 24y - 16xy )  ]  +
       {{F_2}^2}[ {{g_V}^2}
           ( -12x + 12{x^2} - 4{x^4} + 12y - 8{x^2}y -
             16{y^2} ) + {{g_A}^2}
           ( 12x \cr
&&+ 12{x^2} - 4{x^4} + 12y - 8{x^2}y -
             16{y^2} )  ]  \}  +
    {C_7}( {C_9^{eff}} + {C_9^{eff*}} )
     {m_b}m_{\Lambda_b}\{ {{F_1}^2}
        [ {g_A}{h_A}
           ( -6x + 6{x^2} - 6y + 6xy ) \cr
&& + {g_V}{h_V}
           ( -6x - 6{x^2} + 6y + 6xy )  ]  +
       {{F_2}^2}[ {g_V}{h_V}
           ( -6x + 6{x^2} + 6y + 6xy - 12{y^2} )  +
          {g_A}{h_A}
           ( -6x - 6{x^2}\cr
&& - 6y + 6xy + 12{y^2} )
           ]  + {F_1}{F_2}
        [ {g_V}{h_V}
           ( 12x - 12y - 12xy + 12{y^2} )  -
          {g_A}{h_A}
           (12x + 12y - 12xy - 12{y^2} )] \}  \cr
&&+ ( {{{C_{10}}}^2} +
       {|C_9^{eff}|^2} ) m_{\Lambda_b}^2
     \{ {{F_1}^2}
        [ {{{h_A}}^2}
           ( 3x - 2{x^2} + 3{x^3} + 3y - 6xy + 3{x^2}y -
             4{y^2} )  + {{{h_V}}^2}
           ( -3x - 2{x^2} - 3{x^3} \cr
&&+ 3y + 6xy +
             3{x^2}y - 4{y^2} )  ]  +
       {F_1}{F_2}
        [ {{{h_A}}^2}
           ( 6x + 2{x^3} + 6y - 12xy + 6{x^2}y -
             12{y^2} + 4x{y^2} )  +
          {{{h_V}}^2} ( 6x \cr
&&+ 2{x^3} - 6y - 12xy - 6{x^2}y +
             12{y^2} + 4x{y^2} )  ]  +
       {{F_2}^2}[ {{{h_A}}^2}
           ( 3x + 2{x^2} + 3{x^3} + 3y - 6xy - {x^2}y -
             8{y^2} \cr
&&+ 4{y^3} )  + {{{h_V}}^2}
           ( -3x + 2{x^2} - 3{x^3} + 3y + 6xy - {x^2}y -
             8{y^2} + 4{y^3} )  ]  \}  \} \cr
B(y)&=&\frac{2}{3} {{{k_0}}^2} r x m_{\Lambda_b}^3\{-{C_7} ({C_9^{eff}}+C_9^{eff*}) m_b m_{\Lambda_b}[6 F_1 F_2
 (g_V h_A - g_A h_V) x +
           F_1^2 (3 g_V h_A - 3 g_A h_V + g_V h_A x  \cr
&& + g_A h_V x) +
           F_2^2 (-3 g_V h_A + 3 g_A h_V + g_V h_A x + g_A h_V x) +
           2 F_2 (F_1 g_V h_A + 3 F_2 g_V h_A + F_1 g_A h_V \cr
&& - 3 F_2 g_A h_V) y] +
      ({C_{10}}^2 + |C_9^{eff}|^2) m_{\Lambda_b}^2 h_A h_V
       [-4 F_1 F_2 x - F_1^2 (3 + x^2) + F_2^2 (3 + x^2) +
         4 (F_1^2 - 2 F_2^2 \cr
&& + F_1 F_2 x) y + 4 F_2^2 y^2] -
      4 {C_7}^2 g_A g_V m_b^2 [-8 F_1 F_2 x + F_2^2 (3 - x^2) + F_1^2 (-3 + x^2) +
           2 (F_1^2 - 5 F_2^2 \cr
&& + 4 F_1 F_2 x) y + 8 F_2^2 y^2)]\}
\end{eqnarray}
where $r=1/(1+x^2-2 y)$, $k_0^2=|V_{bt}V_{st}|^2 (\alpha G_F /\pi)^2\sqrt{y^2-x^2}/(128\pi^3)$.
\begin{table}
\caption{Wilson coefficients $C_i(\mu)$ for
$\Lambda^{(5)}_{\overline {MS}}=225 \; MeV$, $\mu =5 \; $ GeV and $m_t=174
\;$ GeV. NDR scheme.
\\}
\begin{tabular}{ccc}
$C_1$ & $-0.243$ & \\
$C_2$ & $1.105$ & \\
$C_3$ & $1.083 \times 10^{-2}$ & \\
$C_4$ & $-2.514 \times 10^{-2}$ & \\
$C_5$ & $7.266 \times 10^{-3}$ & \\
$C_6$ & $-3.063 \times 10^{-2}$ & \\
$C_7$ & $-0.312$ & \\
$C_9$ & $4.193 $ & \\
$C_{10}$ & $-4.578$ & \\
\end{tabular}

\end{table}

\vspace*{1.5cm}

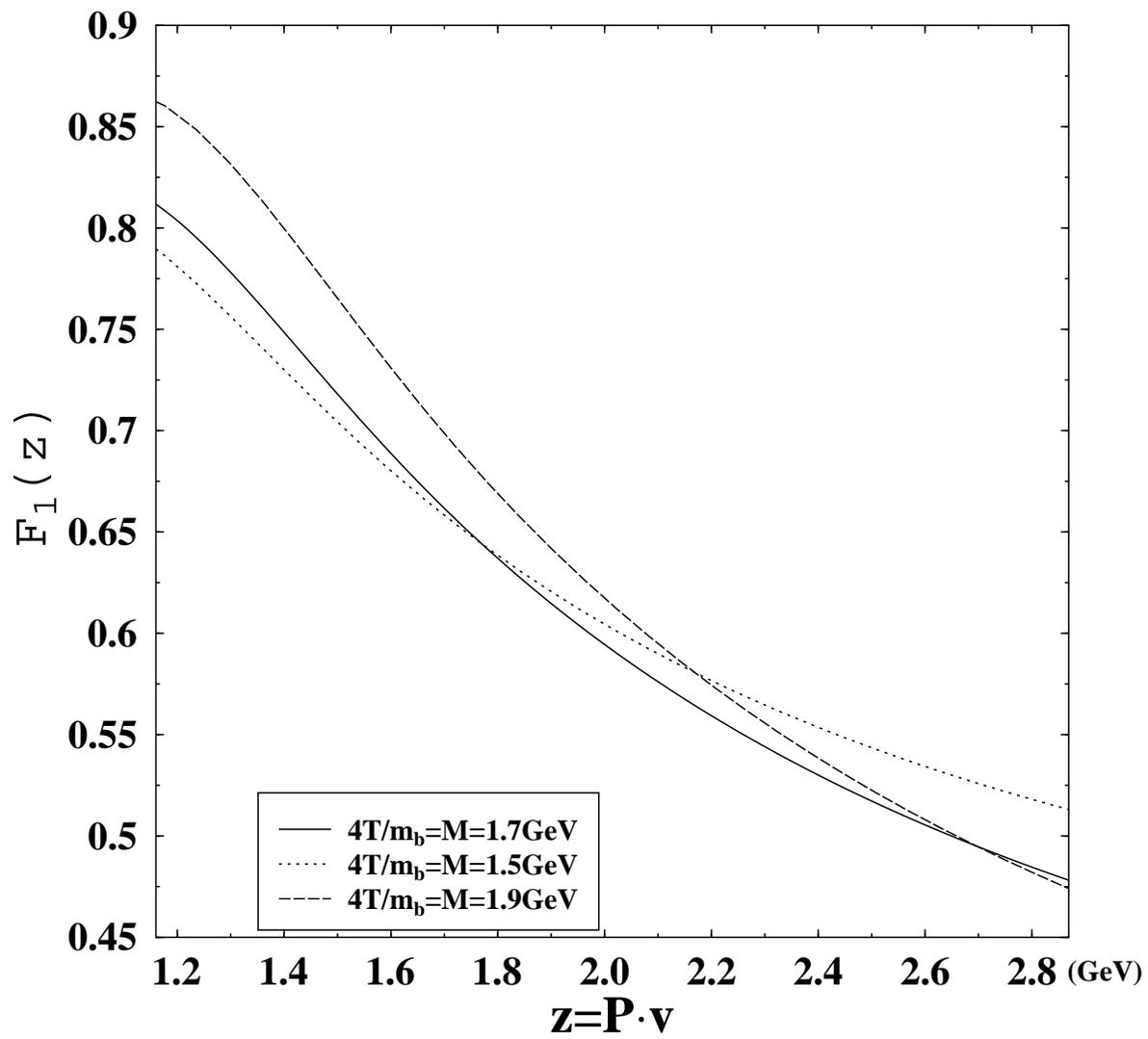
\begin{figure}
\caption{Form factor $F_1(z)$ with different values of Borel parameter
T and M.}
\end{figure}
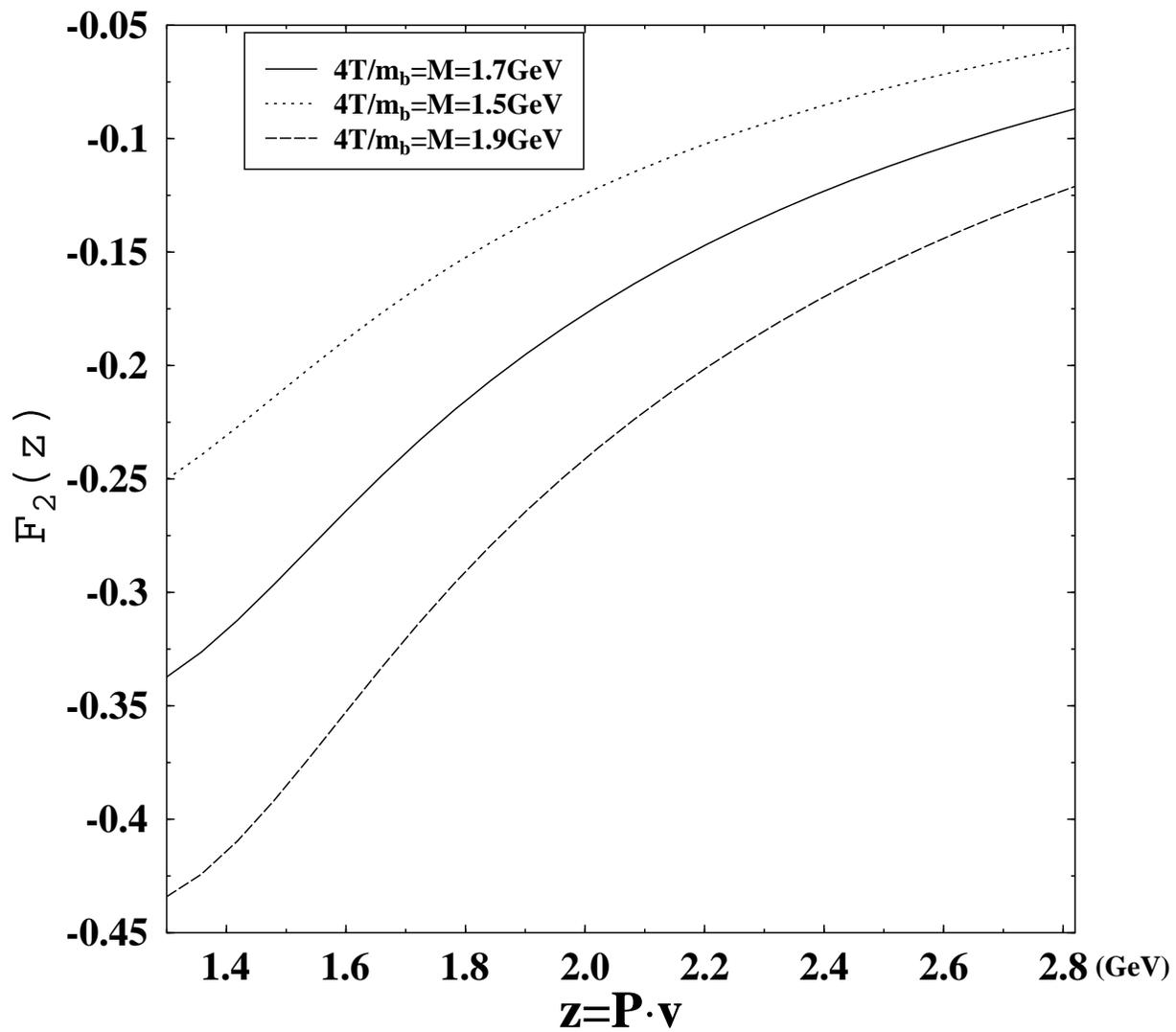
\begin{figure}
\caption{Form factor $F_2(z)$ with different values of Borel parameter
T and M. }
\label{f2}
\end{figure}
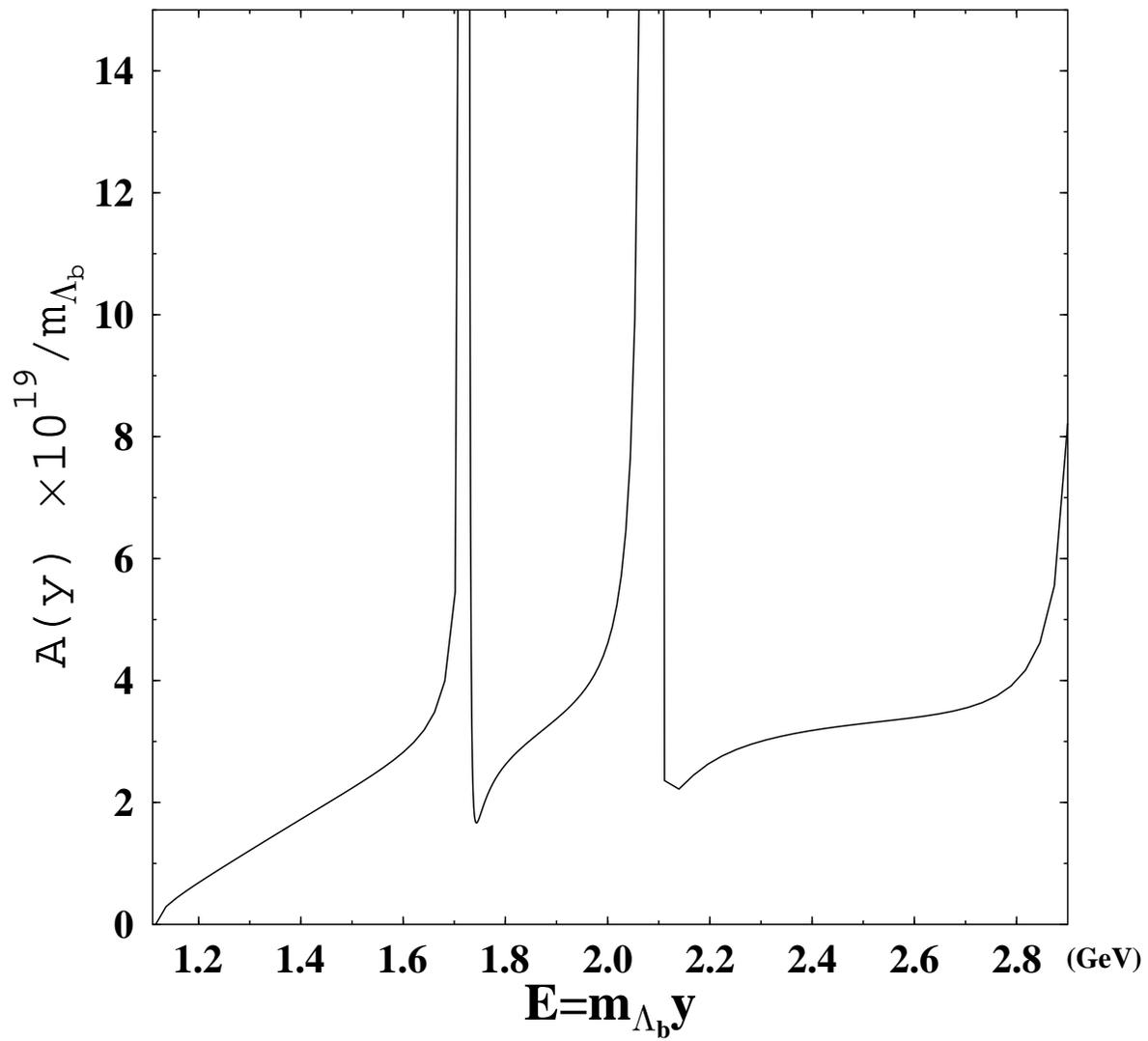
\begin{figure}
\caption{The spin-independent term of the spectrum: A(y)}
\end{figure}
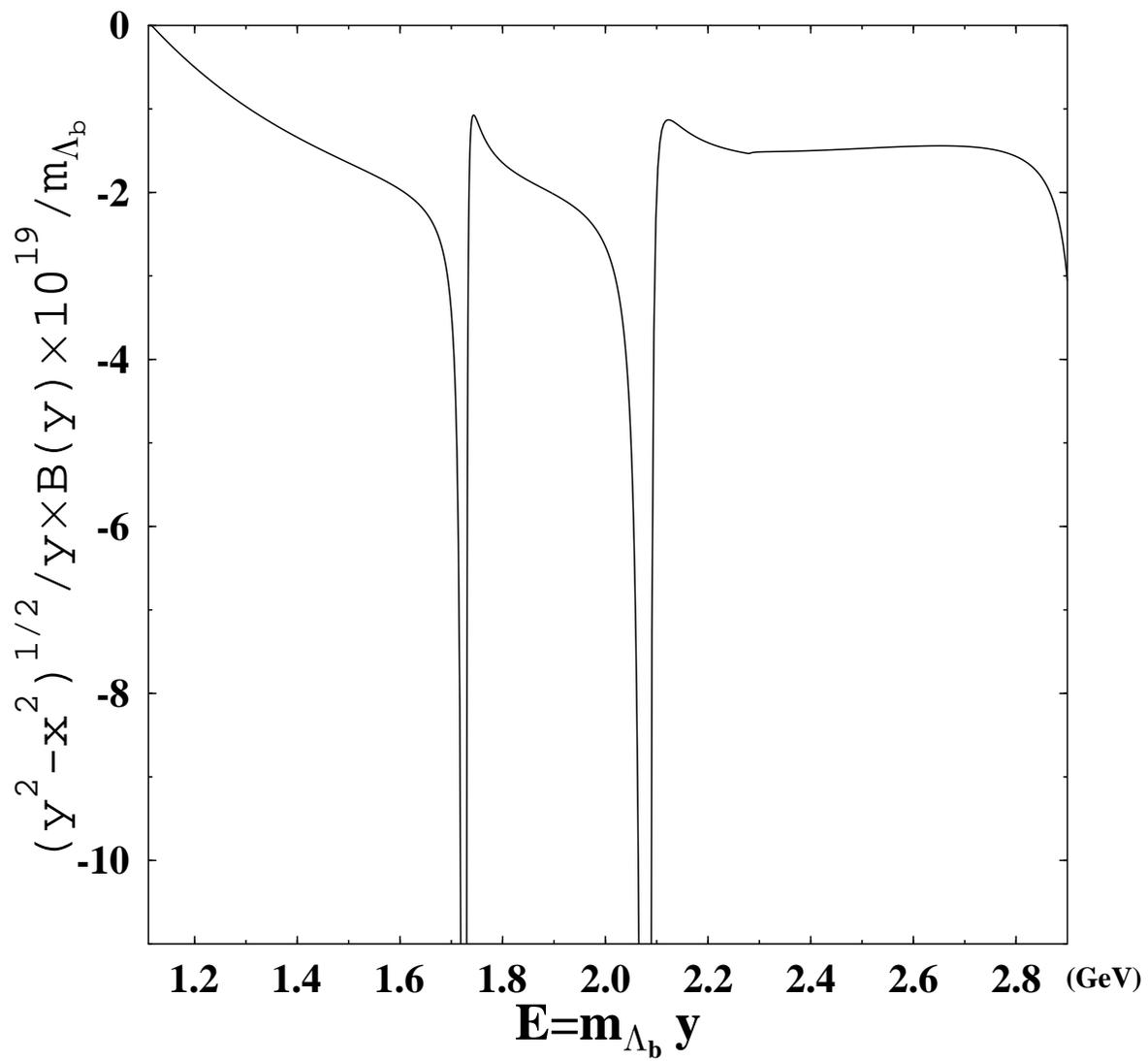
\begin{figure}
\caption{The spin-dependent term of the spectrum: ${\sqrt{y^2-x^2} \over {y}} B(y)$}
\end{figure}

\end{document}